\documentclass[aps,prb,superscriptaddress,showkeys]{revtex4-2}
\usepackage[utf8]{inputenc}
\usepackage{amsmath,amssymb,amsfonts}
\usepackage{multirow}
\usepackage{graphicx,float}
\usepackage[colorlinks=true,
citecolor=blue,
linkcolor=blue,
urlcolor=blue]{hyperref}
\usepackage{footmisc}
\usepackage{array}
\usepackage{array}
\newcolumntype{P}[1]{>{\centering\arraybackslash}p{#1}}

\begin{document}
	
\title{Anomalous Tensile Strain Induced Enhancement in the Lattice Thermal Transport of Monolayer ZnO: A First Principles Study}
\author{Saumen Chaudhuri}
\affiliation{Department of Physics, Indian Institute of Technology Kharagpur, Kharagpur 721302, India}
\author{Amrita Bhattacharya}
\email[corresponding author: ]{b\_amrita@iitb.ac.in}
\affiliation{Department of Metallurgical Engineering and Materials Science, Indian Institute of Technology Bombay, Mumbai 400076, India}
\author{{A. K. Das}}
\affiliation{Department of Physics, Indian Institute of Technology Kharagpur, Kharagpur 721302, India}
\author{{G. P. Das}}
\email[corresponding author: ]{gourpdas@gmail.com}
\affiliation{Research Institute for Sustainable Energy, TCG Centres for Research and Education in Science and Technology, Sector V, Salt Lake, Kolkata 700091, India}
\author{B. N. Dev}
\email[corresponding author: ]{bhupen.dev@gmail.com}
\affiliation{Department of Physics and School of Nano Science and Technology, Indian Institute of Technology Kharagpur, Kharagpur, 721302, India}
\affiliation{Centre for Quantum Engineering, Research and Education, TCG Centres for Research and Education in Science and Technology, Sector V, Salt Lake, Kolkata 700091, India}
	
\begin{abstract}
Density functional theory based calculations have been performed for solving the phonon Boltzmann transport equation to investigate the thermal transport properties of monolayer (ML) ZnO under in-plane isotropic biaxial tensile strain. The in-plane lattice thermal conductivity ($\kappa_{\text{L}}$) of ML-ZnO increases dramatically in response to the biaxial tensile strain ranging from 0\% to 10\%, conflicting the general belief. The strain-induced stiffening of the ZA phonon mode and the resulting concomitant
increase in group velocity and decrease in phonon population is found to play a significant role behind the unusual enhancement of $\kappa_{\text{L}}$. The mode resolved analysis shows the tensile strain driven competitive behavior between different phonon properties, mainly the group velocity and phonon lifetimes, being responsible for the observed unusual enhancement in $\kappa_{\text{L}}$. Additionally, the phonon scattering calculations show the importance of inclusion of 4-phonon scattering in the thermal transport calculations suggesting the significance of higher-order anharmonicity in ML-ZnO. A strikingly high 4-phonon scattering strength in ML-ZnO primarily results from the strong anharmonicity, quadratic ZA mode dispersion, large frequency gap in phonon dispersion, and reflection symmetry induced selection rule. The incorporation of 4-phonon scattering significantly alters the transport characteristics of all the phonon modes, in general and ZA phonons, in particular. At large strains, a linear dispersion of the ZA mode and closure of the frequency gap is observed, which results in significant reduction of 4-phonon scattering strength in ML-ZnO.    
\end{abstract}

\date{\today}
\maketitle
	
\section{Introduction}
The discovery of graphene \cite{novoselov2004electric} and its unique physical and chemical properties have drawn significant attention in recent years towards the atomically thin layered materials \cite{huang2011graphene, srivastava2012functionalized, yoo2011ultrathin, lin2009operation}. Within the ever growing family of two-dimensional (2D) materials, the ones having sizable band gap have attracted tremendous interest due to their potential in electronic and optical applications \cite{wilson1969transition, benameur2011visibility, radisavljevic2011single, pradhan2013intrinsic, radisavljevic2011integrated}. Recent advances in materials engineering have shown that, bulk ZnO which crystallizes in wurtzite structure can be exfoliated into a 2D planar graphene-like stable monolayer (ML) form \cite{claeyssens2005growth, topsakal2009first, tusche2007observation}. The single-layer counterpart of ZnO, with a wide band gap of $\sim$ 3 eV, has emerged as a promising material for electronic and optoelectronic applications. The electrical transport properties of ML-ZnO have been investigated thoroughly over the years \cite{topsakal2009first, kan2013stability, law2005nanowire, hwang2007zno}, while the study on its thermal transport properties is limited. The integration of ML-ZnO in device applications, following its superior electrical performances, makes it imperative to explore the thermal transport properties of ML-ZnO, especially the lattice thermal conductivity ($\kappa_{\text{L}}$). 

Due to the significant challenges associated with the fabrication of high quality 2D materials and the precise measurement of their thermal conductivity, first principles based theoretical approaches, such as density functional theory (DFT) and molecular dynamics (MD) simulations, have become indispensable to predict and analyze the thermal conductivity of these material systems. The $\kappa_{\text{L}}$ of ML-ZnO, using first-principle DFT based calculations, is found to be only 4.5 Wm$^{-1}$K$^{-1}$ \cite{wang2017low}, which is considerably lower compared to its bulk counterpart (48 Wm$^{-1}$K$^{-1}$) and substantially lower compared to its analogous 2D planar systems such as graphene ($\sim$ 3000 Wm$^{-1}$K$^{-1}$) and hexagonal Boron Nitride (h-BN) ($\sim$ 250 Wm$^{-1}$K$^{-1}$) \cite{wang2017low, qin2017orbitally}. Such a low $\kappa_{\text{L}}$ makes ML-ZnO a potential candidate for nanostructure thermoelectrics. However, to possibly explore the superior electrical performances of ML-ZnO, the exclusive operating conditions of modern electronic devices demand a significantly high $\kappa_{\text{L}}$ to dissipate the generated heat. Furthermore, the tunability of thermal transport properties, exhibited for example in h-BN \cite{li2017thermal}, silicine \cite{ding2021anomalous}, or various TMDCs \cite{chaudhuri2023strain, wang2021improved, jia2022high, kuang2015unusual}, could open up avenues for ML-ZnO in numerous electrical and thermal engineering applications.

Strain engineering, due to its easy implementation and reversible application, has emerged as an efficient way to tune the electrical, mechanical, or thermal properties of low-dimensional materials \cite{chaudhuri2023strain, chaudhuri2023strain1, chaudhuri2023ab, chaudhuri2023hydrostatic, das2014microscopic, guzman2014role}. The electrical and vibrational properties of ML-ZnO are seen to be effectively tuned with the application of in-plane mechanical strain \cite{chaudhuri2023ab}. The application of tensile strain, in general, lead to the deterioration of the thermal transport properties of any bulk material. For example, the thermal conductivities of bulk diamond and silicon are found to decrease with tensile strain, owing to the decrease in phonon lifetimes and group velocities \cite{parrish2014origins, li2010strain}. According to the conventional law of phonon transport, when tensile strain is applied, the bonds stretch, resulting in a weakening of the inter-atomic bonds and thereby, the ability of heat conduction should decrease. However, in recent times, in some specific 2D materials an opposite trend i.e., enhancement in $\kappa_{\text{L}}$ with tensile strain, is seen. For example, in bi-layer graphene (BLG) and monolayer C$_{3}$N a non-monotonic up-and-down behavior with tensile strain, i.e. increase at low strain and ultimately decrease at higher strain, is seen \cite{kuang2015unusual, taheri2020highly}. A competition between various phonon properties such as the group velocity and phonon lifetime is found to be responsible behind the non-monotonic variation in $\kappa_{\text{L}}$. In planar h-BN sheet and low-buckled silicine, on the other hand, a continuous increase in $\kappa_{\text{L}}$ with increasing tensile strain is observed \cite{li2017thermal, ding2021anomalous}. The out-of-plane acoustic or the ZA mode phonons are found to play a significant role behind the anomalous increase in $\kappa_{\text{L}}$ with tensile strain. In many other 2D materials, such as monolayer graphene (MLG), various transition metal di-chalcogenides (TMDCs), phosphorene etc. the ZA mode is seen to be the primary heat conduction carrier, owing to its unique quadratic dispersion. However, assuming this unusual increase in $\kappa_{\text{L}}$ with tensile strain as a general trend for any 2D layered material would be misleading. As in numerous 2D materials having analogous crystal symmetry viz. MLG, various TMDCs such as MoS$_{2}$ \cite{chaudhuri2023strain}, WS$_{2}$ \cite{bera2019strain}, HfS$_{2}$ \cite{wang2021improved} etc. the $\kappa_{\text{L}}$ decreases with application of tensile strain. It is, therefore, interesting to investigate the variation in $\kappa_{\text{L}}$ of ML-ZnO with tensile strain, which has similar crystal structure and quadratic dispersion of the ZA phonon branch, like the other 2D materials discussed so far. An increase in $\kappa_{\text{L}}$ of ML-ZnO with strain would become beneficial for its applications as a heat dissipative substrate in nanostructure electronic devices, which is hindered due to its intrinsically low $\kappa_{\text{L}}$. A further decrease in $\kappa_{\text{L}}$, on the other hand, would improve its thermoelectric performance.

Over the years, the first-principles DFT based calculations have been successfully employed in accurately predicting the $\kappa_{\text{L}}$ of various materials at a low computational expense \cite{broido2007intrinsic, garg2011role, li2012thermal, li2012thermal1}. To limit the computational cost, these first-principles calculations consider only the intrinsic 3-phonon scattering term, neglecting the effect of all the higher-order perturbative terms in the crystal Hamiltonian. However, this often is the source of error in the estimation, as first principles calculations, considering only the lowest-order perturbative term overestimates the $\kappa_{\text{L}}$ in a number of materials \cite{feng2016quantum, feng2017four, yang2019stronger, zhang2022four} including single-layer graphene \cite{feng2018four} and MoS2 \cite{gandi2016thermal}. Ruan \textit{et al.} have shown that the subsequent incorporation of the next higher-order perturbative term, which includes the 4-phonon scattering lifetimes, brings down the values of $\kappa_{\text{L}}$ close to the experimentally measured ones \cite{feng2016quantum, feng2017four, yang2019stronger, zhang2022four, feng2018four}. In graphene, the thermal transport calculations considering only the 3-phonon scattering is found to remarkably overestimate the relative contribution of the ZA phonons \cite{feng2018four}. In the low-dimensional materials, such as graphene \cite{feng2018four}, MoS$_{2}$, TaS$_{2}$ \cite{zhang2022four}, diamane \cite{zhu2019giant} etc. the primary reasons behind the strikingly high 4-phonon scattering strength are identified to be (1) strong anharmonicity, (2) large frequency gap in phonon dispersion, (3) mirror reflection symmetry imposed selection rule and (4) the quadratic dispersion of the ZA phonon branch. Since ML-ZnO possesses all these characteristics in its phonon dispersion \cite{wang2017low, chaudhuri2023ab}, it is legitimate to expect a strong 4-phonon scattering in ML-ZnO as well. In other words, the calculations of thermal transport properties of ML-ZnO considering 3-phonon scattering as the only resistive source to phonon transport, is expected to lead to an overestimation of the thermal conductivity and related properties. It is, therefore, instructive to investigate the remarkable impact of the 4-phonon scattering on the thermal transport properties of ML-ZnO. 

ML-ZnO has been attracting an intensive interest from the scientific community owing to its distinctive electronic properties and therefore, have been explored thoroughly over the years. However, in spite of the critical phonon properties that may give rise to unconventional heat conduction behavior, the phonon transport properties of ML-ZnO have not been avidly explored. Here, in this work, we have proposed a strategy to engineer the phonon transport properties and thereby, tune the $\kappa_{\text{L}}$ of ML-ZnO in a wide range. Additionally, we have illustrated the underlying mechanism of the atypical thermal transport behavior of ML-ZnO with tensile strain through systematic first-principles investigations. Moreover, in spite of possessing all the signatures that may lead to a very strong 4-phonon scattering, the influence of higher-order anharmonicity on the thermal transport properties of ML-ZnO has not been studied yet. In this study, we have explored the impact of 4-phonon scattering on the thermal conductivity of ML-ZnO and its variation with application of tensile strain.

\section{Computational details}
First-principles calculations coupled with the phonon Boltzmann transport equation (BTE) have been performed using ab-initio density functional theory (DFT) as implemented in the Vienna Ab Initio Simulation Package (VASP) \cite{kresse1996efficient, kresse1996efficiency} together with projector augmented wave (PAW) potentials to account for the electron-ion interactions \cite{kresse1999ultrasoft}. The electronic exchange and correlation (XC) interactions are addressed within the LDA+U method with a Hubbard-U of 8 eV for the Zn-$d$ states \cite{kohn1965self}. The choice on the exchange-correlation (XC) functional, used in all the calculations, is made after performing a series of DFT+U calculations with different XC functional and U values on the geometrical and electronic properties of ML-ZnO. The lattice parameter, electronic properties and the relative position of the Zn-$d$ states obtained from hybrid functional calculations using HSE06 functional, and experimental results available in literature, are used as a reference behind the determination of the appropriate XC functional and U value. Details of the calculations and results of the same can be found in our earlier study \cite{chaudhuri2023ab}. In all calculations, the Brillouin zone (BZ) is sampled using a well-converged Monkhorst-Pack \cite{monkhorst1976special} k-point set of $ 21\times 21\times 1$, and a conjugate gradient scheme is employed to optimize the geometries until the forces on each atom are less than 0.01 eV/$\text{\AA}$. In the present calculations, in-plane equi-biaxial strain has been implemented by isotropically changing the in-plane lattice parameters to a desired value and then, the internal forces acting on each atom is minimized. The applied strain is defined as $\epsilon = \dfrac{a-a_{0}}{a_{0}} \times 100\%$, where $a_{0}$ and $a$ are the unstrained and the strained lattice parameters, respectively. A vacuum thickness of approximately 20 $\text{\AA}$ is used to avoid the spurious interaction between the periodic images of the layers. 

The phonon dispersion curves are calculated based on the supercell approach using the finite displacement method with an amplitude of 0.015 $\text{\AA}$ as implemented in the phonopy code \cite{togo2015first}. To compute the thermal transport properties, the linear Boltzmann transport equation (BTE) for phonons is solved as implemented in the ShengBTE code \cite{li2014shengbte}. All orders (second, third and fourth) of anharmonic interatomic force constants (IFC) are calculated based on the finite-difference supercell method. For the third-order IFC a $ 6\times 6\times 1$ supercell is created and interactions up to the 6th nearest-neighbour (NN) atom is considered using the THIRDORDER.PY extension module of ShengBTE \cite{li2014shengbte}. Similarly, a $ 4\times 4\times 1$ supercell along with the interactions up to the 3rd NN atoms are considered to compute the fourth-order IFC. A strict energy convergence criterion of $10^{-8}$ eV along with well-converged k-meshes are used in all the supercell based calculations. The scattering rates corresponding to the three- and four-phonon processes have been computed using ShengBTE \cite{li2014shengbte}. To accurately compute the phonon thermal properties, such as the thermal conductivity, scattering phase space and scattering rate, a dense wave-vector mesh or q-mesh of $ 121\times 121\times 1$ is used to sample the reciprocal space of the primitive cells. Note that, for the unstrained and all the strained cases the convergence of the computed thermal conductivity w.r.t. the q-mesh density, NN cutoff is ensured.

Combining the first-principle calculations with the semi-classical Boltzmann transport equation (BTE), the lattice thermal conductivity ($\kappa_{\text{L}}$) of a material can be calculated as:\\

$ \kappa_{\text{L}}^{\text{xy}} = \dfrac{1}{\text{N}_{\text{q}}V} \sum {\hbar}\omega_\lambda {\dfrac{\partial n_{\lambda}^{0}}{\partial T}} \nu_{\lambda , x} \nu_{\lambda , y} \tau_{\lambda} = \dfrac{1}{\text{N}_{\text{q}}V} \sum \text{C}_{\lambda} \nu_{\lambda , x} \nu_{\lambda , y} \tau_{\lambda}$ \\

where $\lambda$ is a particular phonon mode with frequency $\omega_\lambda$ at a particular wave-vector q, V is the volume of the Brillouin zone (BZ), N$_{\text{q}}$ corresponds to the total number of q-points sampled in the first BZ, n$_{\lambda}^{0}$ is the Bose-Einstein distribution function associated with the phonon mode $\lambda$ at a particular temperature T, $\nu_{\lambda , x}$ is the x component of the phonon group velocity, $\tau_{\lambda}$ is the carrier lifetime of the $\lambda$ mode phonons. The term $\hbar \omega_\lambda \dfrac{\partial n_{\lambda}^{0}}{\partial T}$ corresponds to the phonon specific heat of the $\lambda$ mode (C$_{\lambda}$). The intrinsic phonon scattering rates ($\tau_{\lambda}^{-1}$) and therefore, the carrier lifetime ($\tau_{\lambda}$) are computed according to the Fermi’s golden rule and the detailed derivation of the formulas can be found elsewhere \cite{feng2016quantum, feng2017four, yang2019stronger, zhang2022four, feng2018four}. In this work, both the three- and four-phonon scattering processes are considered in the scattering rate calculations. Thus, the total scattering rate corresponding to the phonon mode $\lambda$ is computed by the Matthiessen’s rule \cite{mahan2000many}, given as: \\

${\dfrac{1}{\tau_{\lambda}}} = \dfrac{1}{\tau_{3, \lambda}} + \dfrac{1}{\tau_{4, \lambda}}$ \\

where $\frac{1}{\tau_{3, \lambda}}$ and $\frac{1}{\tau_{4, \lambda}}$ denote the 3-phonon and 4-phonon scattering rates, respectively. The total scattering rate is, therefore, computed by summing the contribution from all the possible phonon modes.

\section{Results and Discussion}
\subsection{3-phonon scattering}

\begin{figure}[h!]
	\centering
	\includegraphics[scale=0.48]{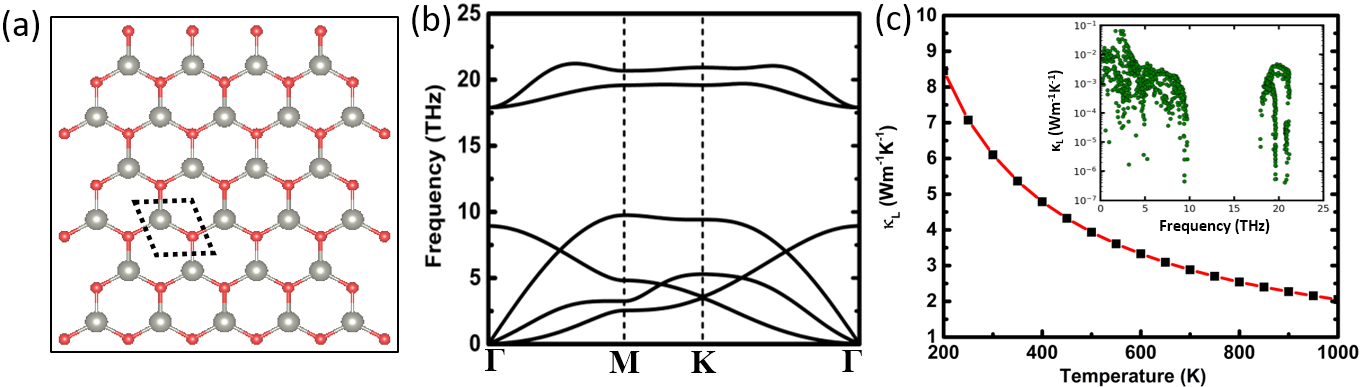}
	\caption{(a) Ball and stick representation of ML-ZnO. The Zn and O atoms are shown in grey and red color respectively. The hexagonal primitive unit cell is depicted using the dashed black box. (b) Phonon dispersion of ML-ZnO plotted along the high symmetry path of the irreducible first Brillouin zone and (c) thermal conductivity ($\kappa_{\text{L}}$) plotted as a function of temperature for the unstrained ML-ZnO. The inset shows the mode-resolved $\kappa_{\text{L}}$ plotted as a function of phonon frequency at a temperature of 300 K.}
	\label{Fig.1}
\end{figure}

The optimized crystal structure of monolayer (ML) ZnO is shown in Fig. \ref{Fig.1}. Like graphene, ML-ZnO adopts a perfectly planar hexagonal honeycomb structural configuration. However, unlike graphene, the primitive unit cell of ML-ZnO consists of two different atoms, one zinc (Zn) and one oxygen (O). The lattice parameter, calculated herein, in the relaxed geometry is found to be consistent with earlier reports using the same exchange and correlation (XC) functional \cite{monteiro2021mechanical, kaewmaraya2014strain}. ML-ZnO, in its pristine form, is found to be a direct band gap semiconductor of band gap value 2.79 eV with both the valence band maximum (VBM) and the conduction band minimum (CBM) at the $\Gamma$-point of the Brillouin zone (BZ) (see Fig. S1).

The phonon band structure of ML-ZnO constructed from the harmonic 2nd order IFCs are shown in Fig. \ref{Fig.1}(b). The absence of imaginary phonon modes suggests the dynamical stability of single-layer ZnO in the graphitic form. Owing to the two atoms in the unit cell the phonon dispersion of ML-ZnO consists of 6 phonon branches, 3 acoustic (ZA, TA and LA) and 3 optical (ZO, TO and LO). Among the 3 acoustic branches, while the TA and the LA mode have linear dispersion, the out-of-plane acoustic mode or the ZA mode is seen to have quadratic dispersion near the BZ center ($\Gamma$-point). Such quadratic dispersion of the ZA mode is understood as a characteristic of any 2D materials, such as graphene \cite{kuang2015unusual}, MoS$_{2}$ \cite{chaudhuri2023strain1} etc. Owing to the large mass ratio of the constituent atoms (M$_\text{Zn}$/M$_\text{O}$ = 4), there exists a large frequency gap in the phonon dispersion of ML-ZnO. The frequency gap of $\sim$ 7.5 THz divides the dispersion into two regions, one low-frequency region consisting of the ZA, TA, LA and the ZO mode and one high-frequency region consisting of the TO and LO mode. Apart from that, one important feature can be immediately noticed from the phonon dispersion of ML-ZnO, e.g. the ZO mode crosses with the low-frequency acoustic modes throughout the BZ, which indicates a strong coupling between the acoustic and the optical phonon modes. These phonon characteristics, such as the quadratic dispersion of the ZA mode, large frequency gap and strong acoustic-optical (a-o) coupling can critically influence the phonon transport properties, as has been discussed later. 

The 3-phonon scattering limited intrinsic lattice thermal conductivity ($\kappa_{\text{L}}$) of ML-ZnO is calculated with temperature in a range of 200 to 1000 K by solving the phonon BTE within the relaxation time approximation (RTA). As can be seen from the $\kappa_{\text{L}}$ vs T plot shown in Fig. \ref{Fig.1}(c), the $\kappa_{\text{L}}$ of ML-ZnO at 300 K is found to be only 6.1 Wm$^{-1}$K$^{-1}$, which agrees well with previous reports. Compared to the other 2D planar materials with the same hexagonal crystal structure, such as AlN (264.1 Wm$^{-1}$K$^{-1}$), hBN (600 Wm$^{-1}$K$^{-1}$) and graphene ($\sim$ 3000 Wm$^{-1}$K$^{-1}$), the $\kappa_{\text{L}}$ of ML-ZnO is significantly low. Such a poor $\kappa_{\text{L}}$ is primarily identified to be resulting from the low group velocity of the acoustic modes and very high intrinsic phonon scattering rates. From the mode resolved plot, shown in Fig. \ref{Fig.1}(c), it is clear that the highest contribution to the $\kappa_{\text{L}}$ originates from the low-frequency phonons in the range of 0-4 THz, albeit mainly ZA phonons. Interestingly, the optical phonons are found to contribute significantly, nearly 30\% of the total $\kappa_{\text{L}}$. Such a large contribution from the optical phonon modes is rarely found, and is partially due to the associated comparatively small scattering rates resulting due to the large frequency gap in the phonon dispersion of ML-ZnO.

Note that, the calculation of $\kappa_{\text{L}}$ of ML-ZnO both in the unstrained and strained conditions have been performed based on the RTA solved phonon BTE. It is understood that the resulting $\kappa_{\text{L}}$ would be underestimated by a magnitude determined by the ratio of the Normal (N) to Umklapp (U) scattering strength. This happens as the RTA consider the N-processes to be a resistive process, while only the U-processes are the source of direct thermal resistance. However, if we assume that the ratio of the N to U scattering strength remains unaltered upon application of strain, then the deviation from the accurate $\kappa_{\text{L}}$ would remain more or less the same at all strain values. Since the focus of the current study is to investigate the modifications in the thermal transport properties of ML-ZnO induced by tensile strain, a small and constant underestimation in $\kappa_{\text{L}}$ would not alter the conclusions. We, therefore, chose to compute the $\kappa_{\text{L}}$ of ML-ZnO for unstrained as well as strained conditions based on the RTA solved phonon BTE. 

\begin{figure}[h!]
	\centering
	\includegraphics[scale=0.4]{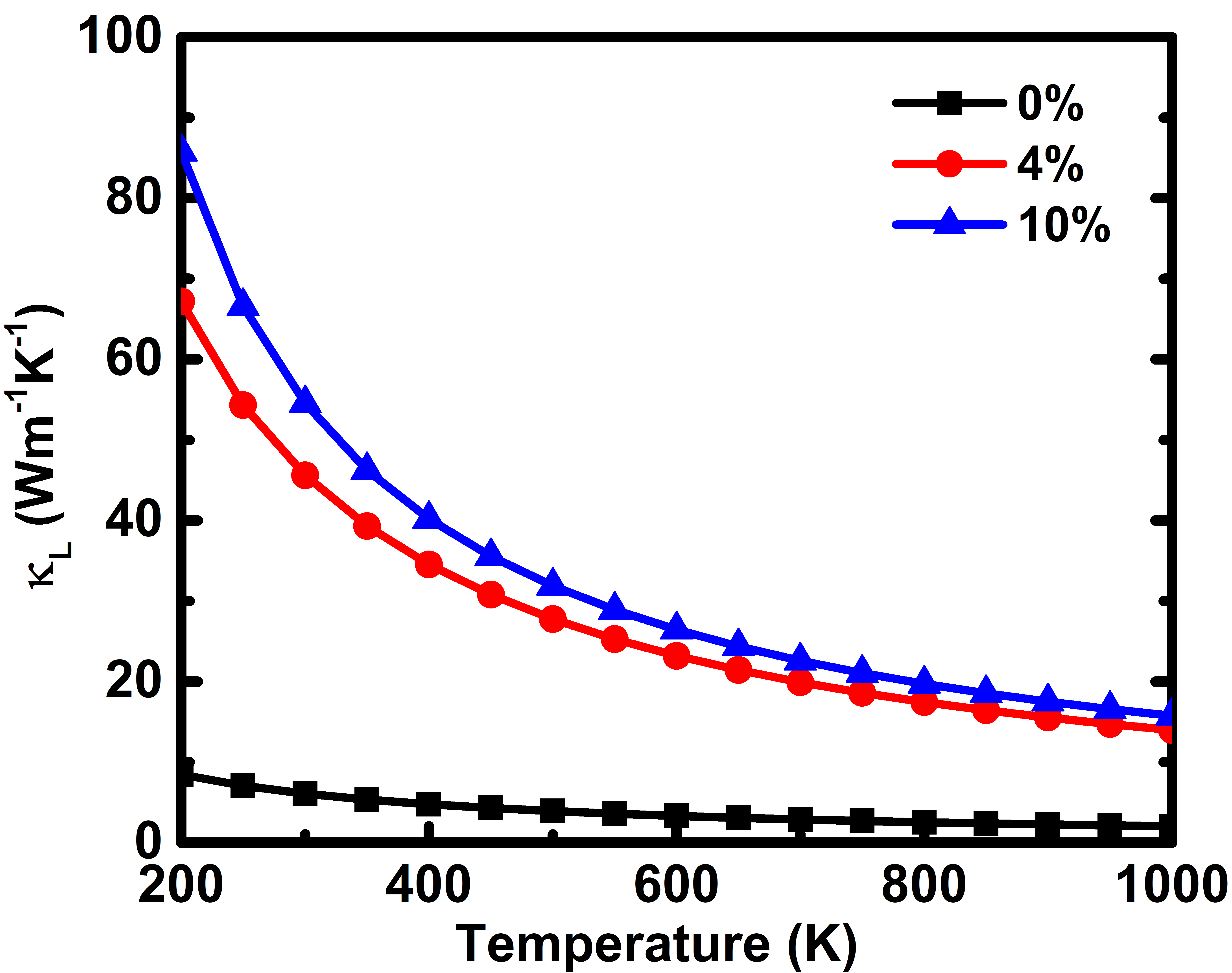}
	\caption{Comparison of temperature-dependent lattice thermal conductivity ($\kappa_{\text{L}}$) of the unstrained ML-ZnO with those of 4\% and 10\% strained conditions.}
	\label{Fig.2}
\end{figure}

To investigate the effect of tensile strain on the thermal transport properties of ML-ZnO, we applied in-plane isotropic biaxial tensile strain of 4\% and 10\% on ML-ZnO. The evolution of $\kappa_{\text{L}}$ of ML-ZnO with the applied strain as a function of temperature is shown in Fig. \ref{Fig.2} and the values of $\kappa_{\text{L}}$ at 300 K are presented in Table \ref{tab:Table 1}. It can be seen that, with the application of tensile strain the $\kappa_{\text{L}}$ increases significantly throughout the temperature range of 200 to 1000 K. This is in stark contrast with the conventional wisdom that the lattice thermal conductivity ($\kappa_{\text{L}}$) of any bulk material reduces with the application of tensile strain \cite{chaudhuri2023strain, wang2021improved, jia2022high, parrish2014origins, li2010strain}. In some low-dimensional materials, however, the $\kappa_{\text{L}}$ is found to behave anomalously with increasing tensile strain; e.g. graphene, C$_{3}$N, silicine and hBN \cite{li2017thermal, ding2021anomalous, kuang2015unusual, taheri2020highly}. For bilayer-graphene \cite{kuang2015unusual} and monolayer C$_{3}$N \cite{taheri2020highly}, the $\kappa_{\text{L}}$ eventually starts to decrease at high values of tensile strain after an initial increase at low strain. Notably, the large decrease in the rate of change of $\kappa_{\text{L}}$ with strain (d$\kappa_{\text{L}}$/d$\epsilon$), for strains in the range of 4\% to 10\%, clearly indicates that the $\kappa_{\text{L}}$ of ML-ZnO may start to decrease, possibly at some larger values of strain ($>$ 10\%). 

\begin{table}[h]
	\caption{\label{tab:Table 1} Room temperature (300 K) lattice thermal conductivity ($\kappa_{\text{L}}$) and specific heat capacity (C$_\text{v}$) of ML-ZnO at diiferent strain levels.}
\centering
\begin{tabular*}{0.5\textwidth}{@{\extracolsep{\fill}}|c|@{\extracolsep{\fill}}c|c|}
	\hline 
	Strain (\%) & $\kappa_{\text{L}}$ (Wm$^{-1}$K$^{-1}$) & C$_\text{v}$ $\times$ 10$^{4}$ Jm$^{-3}$K\tabularnewline
	\hline 
	\hline 
	0 & 6.1 & 31.25\tabularnewline
	\hline 
	4 & 45.6 & 30.20\tabularnewline
	\hline 
	10 & 54.5 & 28.80\tabularnewline
	\hline 
\end{tabular*}
\end{table}

\begin{figure}[h!]
	\centering
	\includegraphics[scale=0.45]{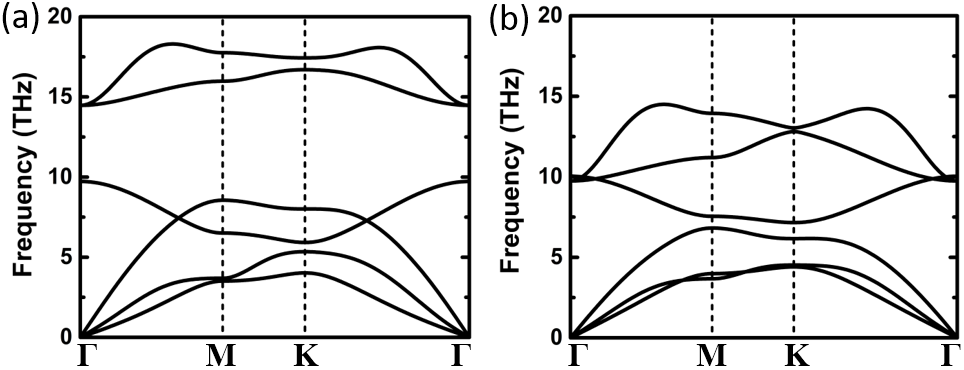}
	\caption{Phonon band structure of ML-ZnO plotted along the high-symmetry path $\Gamma$-M-K-$\Gamma$ under (a) 4\% and (b) 10\% biaxial tensile strain.}
	\label{Fig.3}
\end{figure}

In order to investigate the underlying mechanism behind the tensile strain induced anomalous improvement of the thermal transport properties of ML-ZnO, we have computed the phonon dispersions in the 4\% and 10\% strained conditions and presented in Fig. \ref{Fig.3}. The application of strain alters the dispersion of the in-plane and out-of-plane acoustic modes differently; while the LA and TA modes soften with increasing strain, the ZA mode stiffen. Thereby, the dispersion of the ZA mode turns linear with increasing tensile strain, which shows quadratic behavior in the unstrained condition. A strong bunching of the low-frequency ZA and TA modes are seen to occur at 10\% tensile strain. The high-frequency optical modes (TO and LO) shifts toward lower frequencies with the application of tensile strain. Thereby, the frequency gap separating the TO and LO modes from the low-frequency acoustic and optical modes decreases and ultimately diminishes at a strain of 10\%. Apart from that, the coupling between the ZO and the acoustic modes weaken with increasing strain and completely decouples at a strain of 10\%. These important modifications in the phonon dispersion, such as the (1) quadratic to linear transition of the dispersion of ZA mode, (2) bunching of the acoustic modes, (3) decoupling of the acoustic and ZO mode, and (4) reduction of the frequency gap are expected to have significant impact on the intrinsic phonon scattering mechanism and therefore, the thermal transport properties of ML-ZnO.

To get further insight, we have computed the strain-induced modifications in the various independent parameters that constitute the analytical modelling of $\kappa_{\text{L}}$ ($\kappa_{\text{L}} = \sum\nolimits_{\lambda} \kappa_{{\text{L}}\lambda} = \sum\nolimits_{\lambda} \text{C}_{\lambda} \nu_{\lambda} \nu_{\lambda} \tau_{\lambda}$), such as the specific heat ($\text{C}_{\lambda}$), group velocity ($\nu_{\lambda}$) and relaxation time ($\tau_{\lambda}$) of a particular phonon mode $\lambda$ at temperature 300 K. From the plots of C$_\text{v}$ (= $\sum\nolimits_{\lambda} \text{C}_{\lambda}$) as a function of phonon frequency, shown in Fig. S2, it is clear that the modifications in C$_\text{v}$ is insignificant and likely to be inconsequential in the strain driven enhancement of $\kappa_{\text{L}}$. To get a quantitative measure of the evolution of C$_\text{v}$ with strain, we have presented the room temperature values of C$_\text{v}$ for different strained conditions in Table \ref{tab:Table 1}. The reduction in C$_\text{v}$ with strain clearly suggests its inadequacy in explaining the anomalous enhancement of $\kappa_{\text{L}}$.

\begin{figure}[h!]
	\centering
	\includegraphics[scale=0.45]{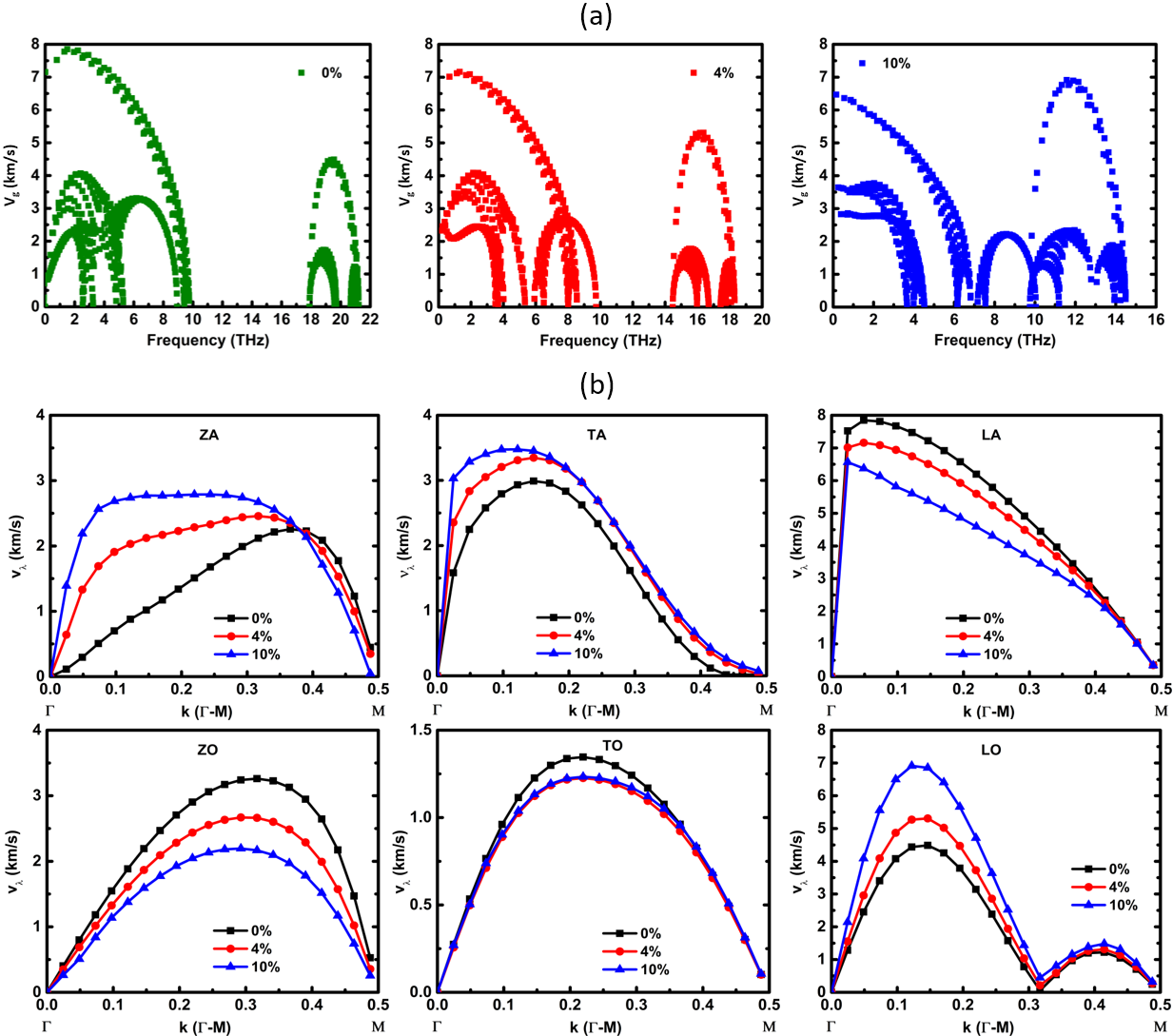}
	\caption{(a) The group velocity ($\nu_{\text{g}}$) of ML-ZnO as a function of phonon frequency for the unstrained (left), 4\% biaxially strained (middle) and 10\% biaxially strained (right) cases. (b) Variation in the mode resolved group velocity ($\nu_{\lambda}$) corresponding to all the individual acoustic and optical phonon modes (i.e. the ZA, TA, LA, ZO, TO and LO modes) w.r.t. the reduced wave vector along the high-symmetry k-path $\Gamma$-M for 0, 4, and 10\% strain. The x-coordinates 0.0 and 0.5 corresponds to the $\Gamma$- and M-point of the Brillouin Zone (BZ), respectively.}
	\label{Fig.4}
\end{figure}

However, the large changes in the phonon dispersion induced by the tensile strain is likely to result in significant modifications in the group velocity of the phonon modes ($\nu_{\text{g}} = \sum\nolimits_{\lambda} \nu_{\lambda}$). The $\nu_{\text{g}}$ of ML-ZnO at 0 K as a function of phonon frequency at different strained conditions are shown in Fig. \ref{Fig.4}(a). Evidently, the highest $\nu_{\text{g}}$ resulting from the LA mode phonons decreases with increasing strain, owing to the weakening of in-plane bonds. The $\nu_{\text{g}}$ corresponding to the high-frequency LO phonons, on the other hand, increases dramatically under the application of tensile strain. In order to elucidate the modifications into greater detail, mode-resolved $\nu_{\text{g}}$ along a specific k-path ($\Gamma$-M) of the BZ is computed, and presented in Fig. \ref{Fig.4}(b). Due to the quadratic to linear transition of the ZA mode dispersion, the corresponding $\nu_{\lambda}$ increases significantly with increasing strain, especially near the BZ center ($\Gamma$-point). At the k-point (0.25, 0, 0), which is midway along the $\Gamma$-M path, the $\nu_{\lambda}$ associated with the ZA phonons increases from 1.67 km/s in the unstrained condition to 2.79 km/s in 10\% strained condition. Among the 6 phonon modes, the $\nu_{\lambda}$ associated with the TA and LO phonons also increase with strain, apart from the ZA mode. The enhancement in $\nu_{\lambda}$ associated with the ZA, TA and LO phonon modes indicate a higher contribution in thermal transport from these modes at the strained conditions.

\begin{figure}[h!]
	\centering
	\includegraphics[scale=0.35]{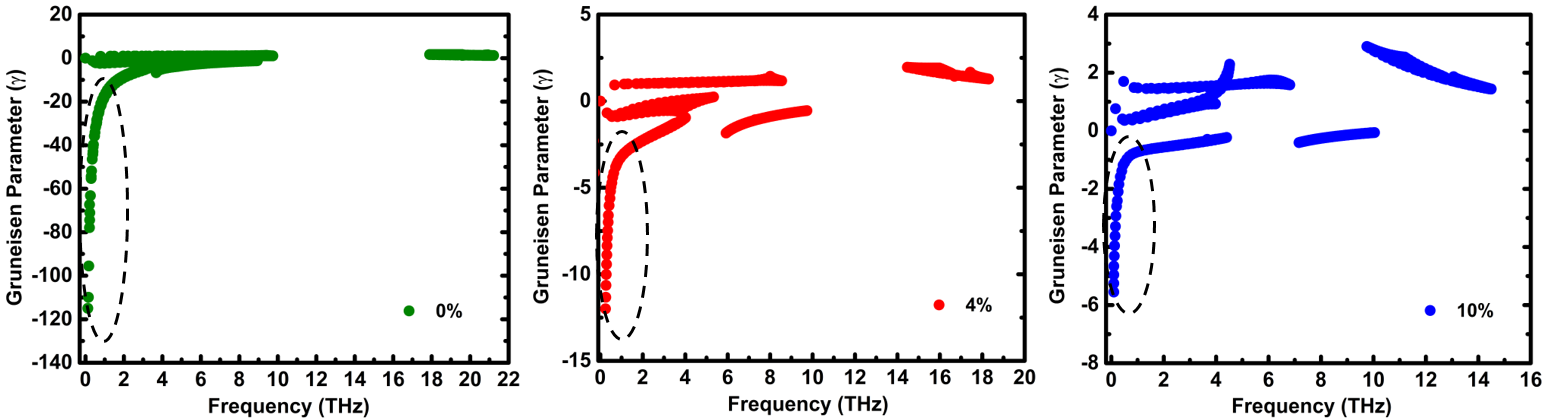}
	\caption{The Gr$\ddot{\text{u}}$neisen parameter ($\gamma$) of ML-ZnO plotted as a function of phonon frequency for the unstrained, and 4\%, 10\% biaxially strained cases. The $\gamma$ corresponding to the ZA mode is highlighted with the black dashed line.}
	\label{Fig.5}
\end{figure}

The most relevant phonon property that governs the transport characteristics of a material is the phonon lifetime ($\tau_{3}$), which is determined from the intrinsic phonon scattering rates ($\tau_{3}^{-1}$) corresponding to the 3-phonon scattering processes resulting from the lowest-order crystal anharmonicity. Presumably, two parameters govern the 3-phonon scattering strength for any material, e.g. (1) Gr$\ddot{\text{u}}$neisen parameter ($\gamma$), (2) 3-phonon scattering phase space (P$_{3}$). To understand the changes in the phonon scattering rates with strain, we first have analyzed the strain induced modifications in these two parameters. The Gr$\ddot{\text{u}}$neisen parameter ($\gamma$) characterizes the crystal anharmonicity of a material and thus, is a measure of the asymmetry present in atomic vibrations. The intrinsic phonon scattering rate is strongly influenced by the crystal anharmonicity. The mode Gr$\ddot{\text{u}}$neisen parameter ($\gamma_\text{i}$) can be defined as $\gamma_\text{i} = - (\text{V}_{0}/\omega_\text{i})(\Delta \omega_\text{i}/\Delta \text{V}_{0})$, where $\omega_\text{i}$ is the frequency of a phonon mode i at the equilibrium volume V$_{0}$. The higher the magnitude of $\gamma_\text{i}$, the larger the anharmonicity, and stronger will be the associated phonon scattering strength. The highest values of $\gamma_\text{i}$ of ML-ZnO, as can be seen from Fig. \ref{Fig.5}, occur in the low-frequency regime (0-2 THz), which belongs to the acoustic phonon modes, albeit mainly the ZA phonons. With increasing tensile strain, the $\lvert \gamma_\text{i} \rvert$ corresponding to the ZA mode decreases significantly, i.e. from $\sim$ 120 in the unstrained condition to $\sim$ 6 in 10\% strained condition. Such a reduction in $\lvert \gamma_\text{i} \rvert$ is expected to result in the decrease of the anharmonic phonon scattering strength associated with the ZA phonons. This fact constitutes an increase in the lifetime ($\tau_{3}$) of the ZA phonons with application of strain. However, the $\lvert \gamma_\text{i} \rvert$ associated with the other phonon modes of ML-ZnO are found to remain unchanged or increase marginally with an increase in tensile strain (see Fig. \ref{Fig.5}).

\begin{figure}[h!]
	\centering
	\includegraphics[scale=0.4]{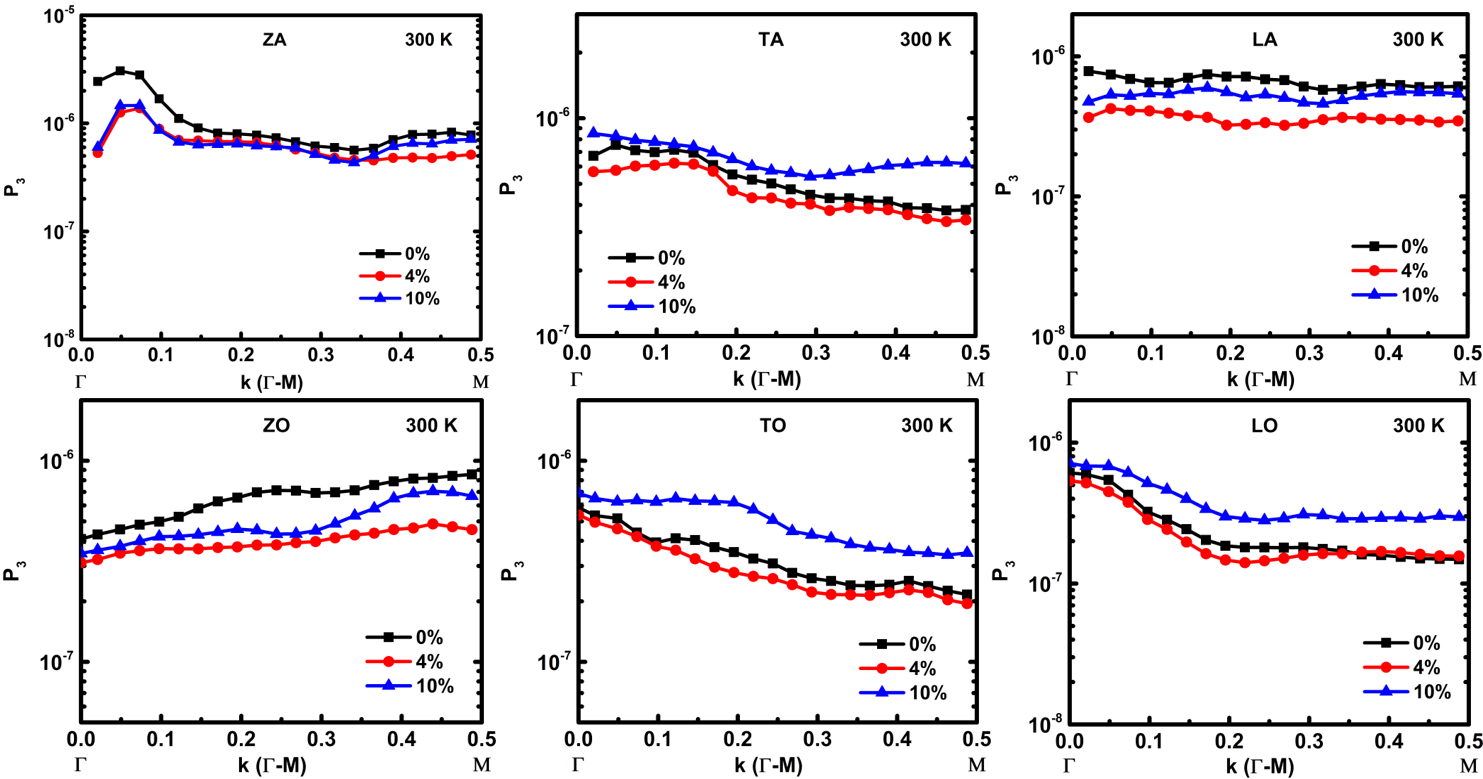}
	\caption{Variation in the mode-resolved 3-phonon phase space (P$_{3}$) of ML-ZnO as a function of the reduced wave vector along the k-path $\Gamma$-M at a temperature of 300 K under the unstrained, and 4\%, 10\% strained conditions for the acoustic ZA, TA and LA (top) and optical ZO, TO and LO (bottom) modes.}
	\label{Fig.6}
\end{figure}

Three-phonon phase space (P$_{3}$), on the other hand, is a measure of the available scattering channels involving three phonons satisfying both the energy and momentum conservation criteria. For a particular phonon mode and for an individual scattering channel, the higher the P$_{3}$, the more probable it is for that particular phonon scattering event to take place. Therefore, the P$_{3}$ has a significant control over the phonon scattering mechanism and the overall thermal transport. However, one should note that the 3-phonon scattering rate ($\tau_{3}^{-1}$) does not depend entirely on the P$_{3}$, but also depends proportionally on the square of the corresponding scattering matrix element. The total P$_{3}$ of ML-ZnO, including both the 3-phonon combination and splitting processes, i.e. the two possible ways three phonons can engage in a scattering event, for different phonon modes along the k-path $\Gamma$-M under strains of 0\%, 4\%, and 10\% are presented in Fig. \ref{Fig.6}. For the unstrained case, the highest values of P$_{3}$ are found to be associated with the ZA mode. Due to the quadratic dispersion of the ZA phonon branch and the resulting high density of low-frequency ZA phonons, the corresponding scattering probability is also very high. However, with the application of strain, the quadratic dispersion turns into a linear one and thereby, the available ZA phonon population decreases, which in turn reduces the corresponding P$_{3}$. For all the 6 phonon modes, the P$_{3}$ decreases for a strain of 4\%, and then increases significantly at 10\% strain. The reduction in P$_{3}$ at 4\% strain is more pronounced for the LA and the ZO mode phonons, due to the strain-induced decoupling of the ZO mode from the acoustic modes. At the strain of 10\%, the frequency gap separating the TO and LO modes from the low-frequency phonon modes diminishes. Thereby, the coupling between the acoustic and the optical modes strengthen and thus, the acoustic phonons can easily scatter into the high-frequency optical modes at a strain of 10\%. As a result, the scattering probability or the P$_{3}$ corresponding to all the phonon modes increase. However, the P$_{3}$ associated with the high-frequency TO and LO mode phonons increase the most. The strain mediated modifications in the P$_{3}$ of the individual phonon modes are likely to be reflected in the phonon scattering rate and thus, the phonon lifetime as well.

\begin{figure}[h!]
	\centering
	\includegraphics[scale=0.5]{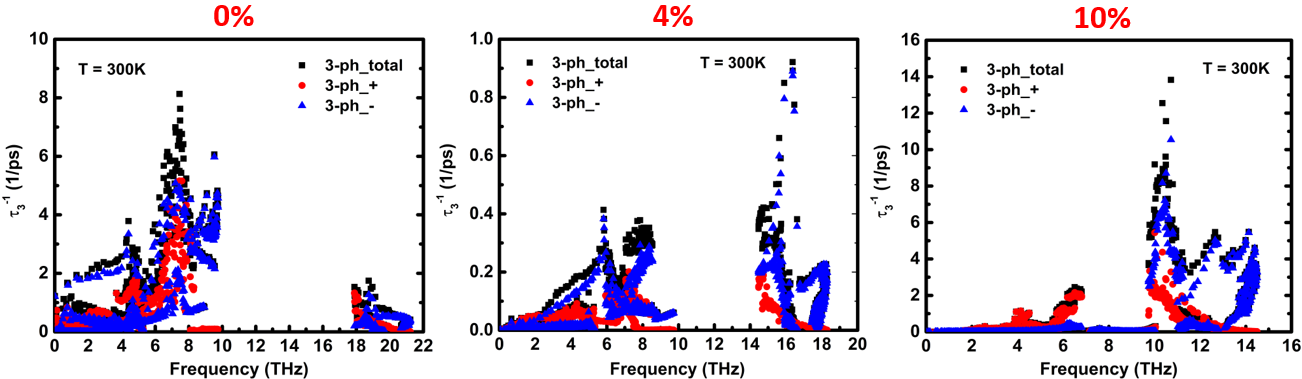}
	\caption{Variation in the 3-phonon scattering rates ($\tau_{3}^{-1}$) of ML-ZnO with phonon frequency at 300 K under unstrained (left), 4\% (middle) and  10\% (right) biaxially strained conditions. The plots include the total 3-phonon scattering rates along with the  separate individual 3-phonon scattering rates corresponding to the  combination (+) and splitting (-) processes. Here, combination refers to the phenomena where two phonons merge to form one phonon and splitting refers to the process where one phonon splits into two new phonons.}
	\label{Fig.7}
\end{figure}

The variation in the overall 3-phonon scattering rate ($\tau_{3}^{-1}$) of ML-ZnO at temperature 300 K for different strained conditions (0\%, 4\% and 10\%) are shown in Fig. \ref{Fig.7}. The $\tau_{3}^{-1}$ of unstrained ML-ZnO is found to be significantly high compared to the other 2D materials with similar crystal structure, such as MoS$_{2}$ \cite{chaudhuri2023strain} or graphene \cite{kuang2015unusual}. Furthermore, the scattering rates corresponding to the acoustic modes are seen to be higher compared to the optical modes, which is rarely found in literature. With the application of tensile strain, the $\tau_{3}^{-1}$ corresponding to both the acoustic and the optical modes change in a non-monotonic manner. At 4\% strain, the $\tau_{3}^{-1}$ drops significantly throughout the frequency range, while increases dramatically at a strain of 10\%. Phonon scattering events involving three different phonons can occur in two possible ways, such as (1) Combination (+): two phonons combine to create a new phonon, and (2) Splitting (-): a single phonon split into two phonons. The sum of the scattering rates corresponding to these two processes constitute the total 3-phonon scattering rate ($\tau_{3}^{-1}$). The scattering rates corresponding to the combination and splitting processes as a function of phonon frequency are also presented in Fig. \ref{Fig.7}. It can be seen that, for the high-frequency phonons the scattering rate corresponding to the splitting process is relatively higher compared to the combination process due to the energy conservation constraint.

\begin{figure}[h!]
	\centering
	\includegraphics[scale=0.4]{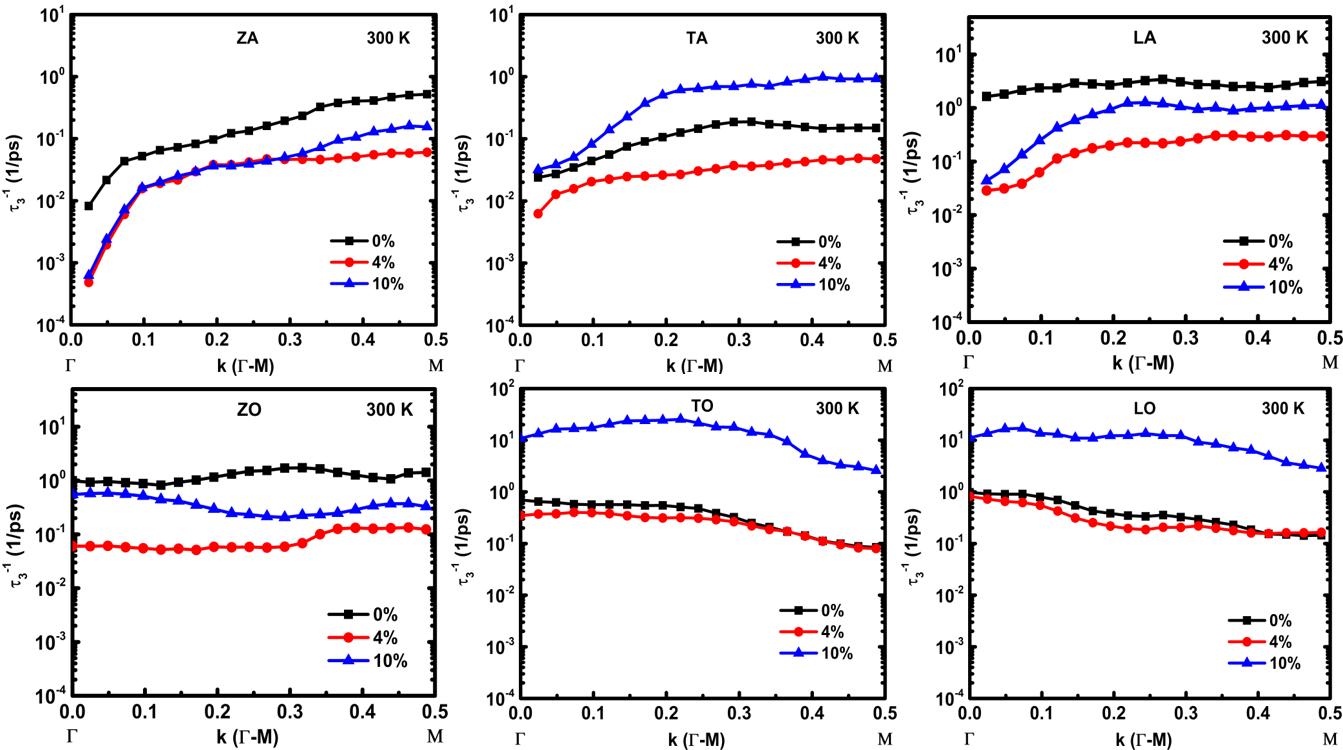}
	\caption{The 3-phonon scattering rates ($\tau_{3}^{-1}$) corresponding to all the phonon modes of ML-ZnO at 300 K for the different strained conditions plotted as a function of the reduced wave vector along the k-path $\Gamma$-M.}
	\label{Fig.8}
\end{figure}

The mode-resolved phonon scattering rates ($\tau_{3}^{-1}$) at 300 K along the k-path $\Gamma$-M for the unstrained and the strained cases are shown in Fig. \ref{Fig.8}. It can be seen that, for pristine ML-ZnO near the zone center ($\Gamma$-point) the $\tau_{3}^{-1}$ is lowest for the ZA mode phonons. With increasing strain, the $\tau_{3}^{-1}$ of the ZA phonons further lowers. The lowest values of $\tau_{3}^{-1}$ for the ZA phonons occur due to two reasons; (1) large frequency gap in the phonon dispersion, (2) reflection symmetry in ML-ZnO. Due to the large frequency gap, the low-frequency ZA phonons can hardly scatter into the high-frequency optical branches obeying the energy conservation criterion. Thereby, the ZA phonons can only scatter into some other low-frequency acoustic phonons and thus, the associated lifetime ($\tau_{3}$) becomes large. Furthermore, the horizontal reflection symmetry in ML-ZnO imposes a unique selection rule for the allowed phonon scattering events, in addition to the energy and momentum conservation rule. According to the selection rule, phonon scattering processes of any order that involve an odd number of out-of-plane phonons (flexural) are forbidden. Therefore, 3-phonon scattering processes may involve only zero or two flexural phonons. Thus, a significant number of scattering channels involving odd numbers of ZA phonons are restricted. This unique selection rule, combined with the large frequency gap, results in a remarkably high lifetime ($\tau_{3}$) corresponding to the ZA phonons. Similar overestimation in the ZA phonon lifetime and their contribution in thermal transport have been observed in graphene \cite{kuang2015unusual} and ML-MoS$_{2}$ \cite{chaudhuri2023strain, chaudhuri2023hydrostatic}. For the strained cases, owing to the reduced Gr$\ddot{\text{u}}$neisen parameter ($\gamma_\text{i}$) and phonon population in the low-frequency regime, the ZA phonon scattering rate drops further. In general, for 4\% tensile strain, the $\tau_{3}^{-1}$ reduces for all the phonon modes throughout the k-path $\Gamma$-M. However, the large enough frequency gap, even at 4\% strain, restricts the TO and LO mode phonons to couple with the low-frequency phonons and thereby, the reduction in $\tau_{3}^{-1}$ for the TO and LO mode is minimal. Except for the ZA mode, since the modifications in the $\gamma_\text{i}$ of the other individual phonon modes are not substantial, the variation in P$_{3}$ alone can well explain the strain induced changes in $\tau_{3}^{-1}$. Owing to the closure of the frequency gap at 10\% strain, while the $\tau_{3}^{-1}$ corresponding to all the phonon modes increase significantly following the increase in P$_{3}$, the $\tau_{3}^{-1}$ of the ZA phonons remains nearly unaltered. The TA mode phonons, on the other hand, having nearly the same dispersion that of ZA mode, experiences a large increase in $\tau_{3}^{-1}$. This can be understood as due to the additional selection rule imposed on the ZA phonons, which restricts a single ZA phonon to couple with two other in-plane phonons or three ZA phonons to couple among themselves. Analyzing the strain induced modifications of the $\tau_{3}^{-1}$ can provide useful insight into the phonon transport mechanism in ML-ZnO and its variation with strain.

\begin{figure}[h!]
	\centering
	\includegraphics[scale=0.35]{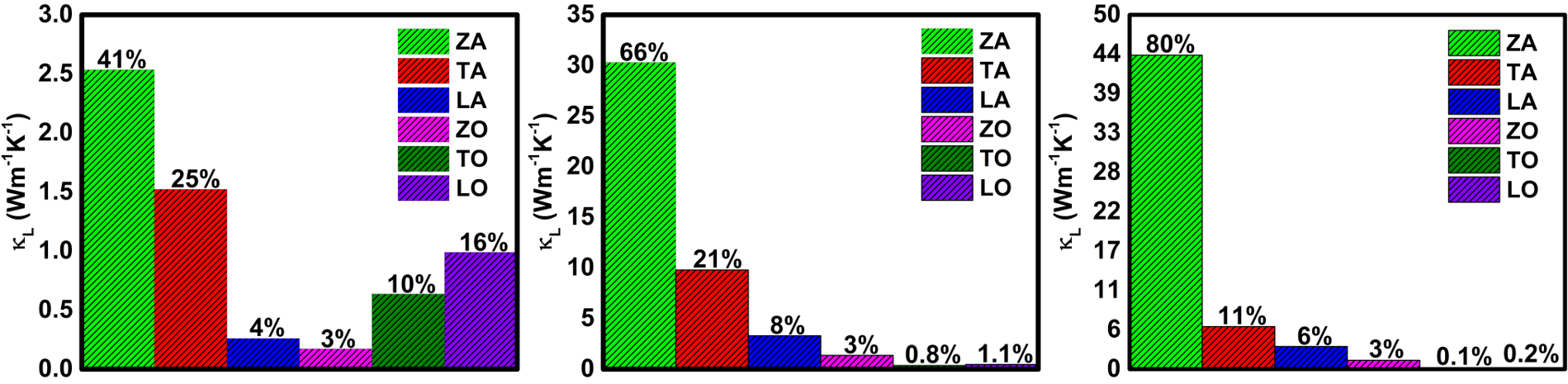}
	\caption{The contribution from the individual phonon modes to the total thermal conductivity ($\kappa_{\text{L}}$) of ML-ZnO at 300 K under the unstrained, 4\% and 10\% biaxial tensile strained condition.}
	\label{Fig.9}
\end{figure}

Combining the effects of strain on the various independent parameters, such as the C$_\lambda$, $\nu_{\lambda}$, and $\tau_{3}$, the underlying physics responsible for the anomalous enhancement in $\kappa_{\text{L}}$ with tensile strain can be constituted. In order to understand the effect of strain on the thermal transport behavior of the individual phonon modes, the mode-resolved contribution to $\kappa_{\text{L}}$ at 300 K for different strain levels are plotted in Fig. \ref{Fig.9}. In the unstrained condition, owing to the restricted phase space and thereby, resulting large phonon lifetime, the ZA phonons carry the majority of heat. With the application of tensile strain, both the $\nu_{\lambda}$, and $\tau_{3}$ corresponding to the ZA phonon increases significantly. Due to the concomitant increase in $\nu_{\lambda}$, and $\tau_{3}$, the contribution from the ZA phonons in the $\kappa_{\text{L}}$ of ML-ZnO increases remarkably with tensile strain. At the 10\% strained condition, the ZA phonons alone carry 80\% of the total heat of ML-ZnO. For all the other phonon modes, a competing behavior between the $\nu_{\lambda}$ and $\tau_{3}$ with strain is seen. Such as, for the TA phonons the $\nu_{\lambda}$ increases and $\tau_{3}$ decreases with strain, whereas for the LA phonons the $\nu_{\lambda}$ decreases and $\tau_{3}$ increases with strain. The compensation between these two effects is responsible for the observed variation in the mode-resolved $\kappa_{\text{L}}$ of ML-ZnO with strain. The application of strain diminishes the thermal transport of the high-frequency TO and LO modes, by reducing their combined contribution from $\sim$ 25\% in the unstrained condition to 0.3\% in the 10\% strained condition. The large increase in $\tau_{3}^{-1}$ corresponding to the TO and LO phonons at 10\% strain results in a very small phonon lifetime, which overshadows the increase in $\nu_{\lambda}$ of the LO phonons. Therefore, in conclusion, all these effects of tensile strain on various phonon parameters, specifically the $\nu_{\lambda}$ and $\tau_{3}^{-1}$, combined together result in the anomalous increasing behavior of $\kappa_{\text{L}}$.

\subsection{3-phonon + 4-phonon scattering}

The strikingly high dominance of the ZA phonons in the thermal transport mechanism of ML-ZnO in the unstrained as well as in the strained conditions poses a serious question towards the competency of the 3-phonon based calculations. Note that, in all the calculations of thermal transport properties of ML-ZnO, discussed so far, the intrinsic phonon scattering rates are calculated considering only the 3-phonon scattering processes, which originates from the first-order anharmonic term in the crystal Hamiltonian. The higher-order anharmonic terms are neglected owing to their presumably small scattering strength. Recently, in some materials including bulk and 2D, the 4-phonon scattering strength is found to be unusually high, even higher than the 3-phonon scattering strength \cite{feng2016quantum, feng2017four, yang2019stronger, zhang2022four, feng2018four}. For example, the inclusion of 4-phonon scattering term in the calculations is found to reduce the relative contribution of the ZA phonons in the thermal transport from 70\% to 30\% \cite{feng2018four}. The phonon properties that are held responsible for the large 4-phonon scattering strength, as discussed in the Introduction section, can be seen to exist in ML-ZnO as well. These features along the remarkably high thermal conduction associated with the ZA phonons lead to the conjecture about a strong 4-phonon scattering in ML-ZnO. It is, therefore, important to examine the extent to which the higher-order anharmonicity can affect the thermal transport characteristics of ML-ZnO, in general, and the ZA phonons, in particular.

\begin{figure}[h!]
	\centering
	\includegraphics[scale=0.5]{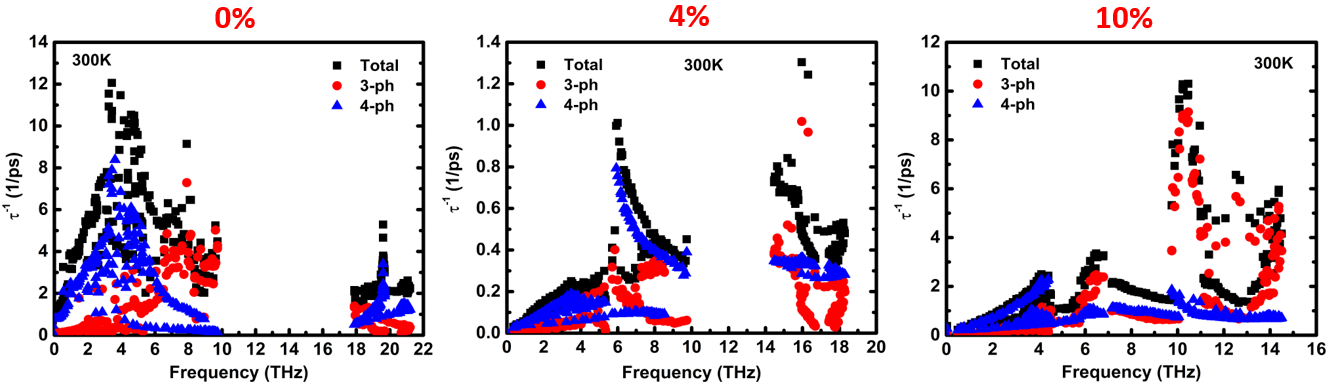}
	\caption{The variation in the 3-phonon (3-ph; red dots), 4-phonon (4-ph; blue dots) and total (3-ph+4-ph; black dots) scattering rates of ML-ZnO  at 300 K plotted as function of frequency for three different strained cases i.e. the unstrained, 4\% strained and the 10\% strained cases.}
	\label{Fig.10}
\end{figure}

To understand the significance of 4-phonon scattering in the thermal transport mechanism of ML-ZnO, we have calculated the 4-phonon scattering rate ($\tau_{4}^{-1}$) under different strained conditions. Therefore, when both the 3-phonon and 4-phonon scattering processes are taken into consideration, the total scattering rate is computed by the Matthiessen’s rule \cite{mahan2000many}, given as: $\tau^{-1} = \tau_{3}^{-1} + \tau_{4}^{-1}$. Note that, compared to the two ways for a 3-phonon process, a 4-phonon scattering process involving four phonons can occur in three ways. Such as, combination (++): three phonons may combine to form a new phonon, splitting (--): a single phonon may split into three phonons, and redistribution (+-): two phonons may combine to form two new phonons. The variation in the total phonon scattering rate ($\tau^{-1}$) of ML-ZnO as a function of phonon frequency at 300 K at different strain levels are presented in Fig. \ref{Fig.10}. At 300 K, the 4-phonon scattering rate is comparable or even higher compared to the 3-phonon scattering rate for the unstrained as well as the strained cases. As the $\tau_{4}^{-1}$ scales quadritically with temperature compared to the linear scaling of the $\tau_{3}^{-1}$ \cite{feng2016quantum}, the 4-phonon scattering dominates the overall phonon scattering mechanism at higher temperatures. This surprisingly high $\tau_{4}^{-1}$ compared to the $\tau_{3}^{-1}$ is against the conventional wisdom regarding intrinsic phonon scattering processes. The general notion following the perturbation theory leads to the fact that, strength of higher-order phonon scattering originates from the corresponding higher-order terms in the crystal Hamiltonian, which gets progressively smaller with the order number. For unstrained ML-ZnO, the $\tau_{4}^{-1}$ is found to be especially dominating in the low-frequency regime e.g. 0-5 THz, which is largely contributed by the low-frequency acoustic branches. Apart from the contribution from the ZA mode at the lowest frequencies, the large $\tau_{4}^{-1}$ around 4 THz originates from the strongly bunched acoustic branches with the ZO branch. The bunching of the phonon modes gives rise to a very high scattering rate corresponding to the 4-phonon redistribution process (+-), as can be seen from Fig. S3. The total scattering rate ($\tau^{-1}$) at 300 K decreases initially at a strain of 4\% then increases dramatically for 10\% strain, which is similar to what is seen with 3-phonon scattering alone. However, the relative strength of 4-phonon scattering ($\tau_{4}^{-1}$/$\tau_{3}^{-1}$) decreases monotonically with increasing strain. Although the reflection symmetry in ML-ZnO and thereby imposed selection rule persists in the strained conditions, the quadratic to linear transformation of the ZA mode dispersion and the abolition of the frequency gap is responsible for the observed drop in the relative strength of 4-phonon scattering. Similar phenomenon has been observed in ML-MoS$_{2}$ as well.

\begin{figure}[h!]
	\centering
	\includegraphics[scale=0.4]{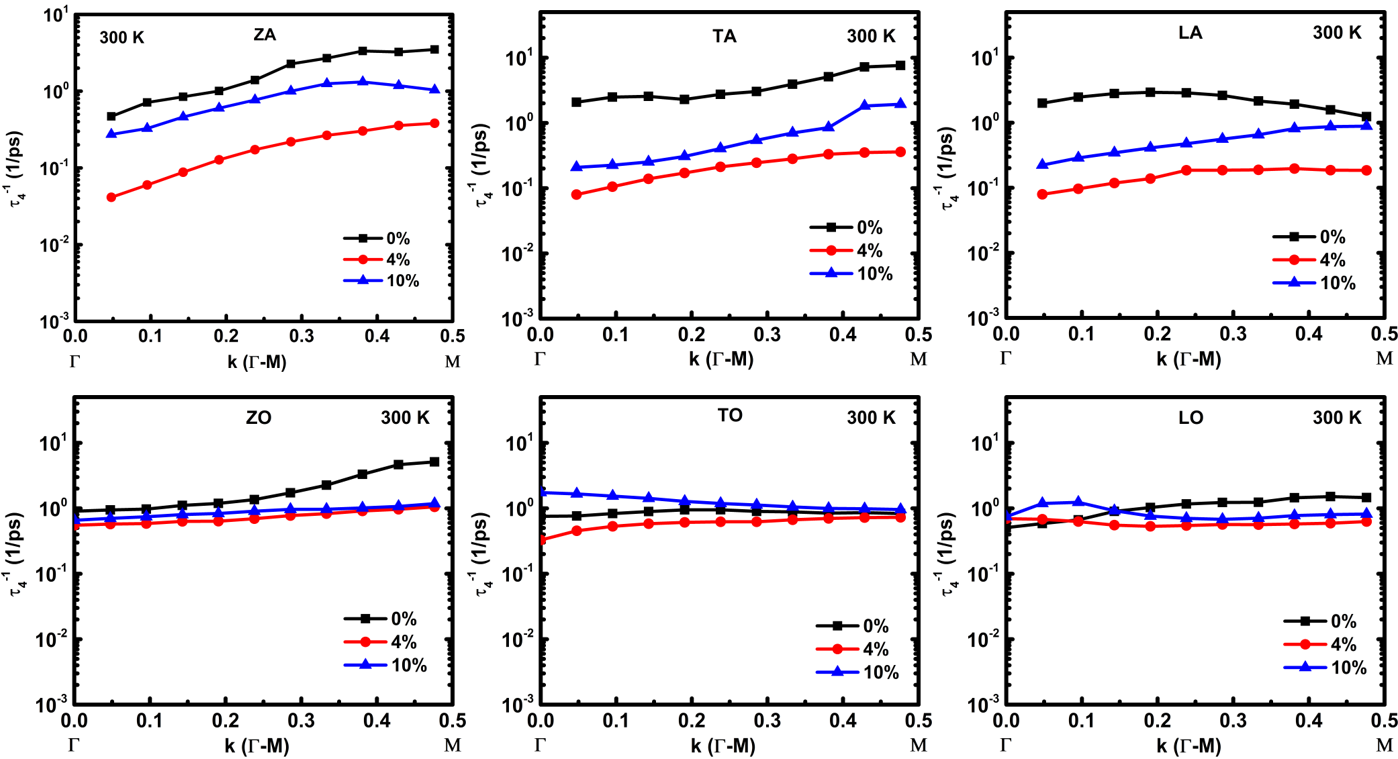}
	\caption{The 4-phonon scattering rates ($\tau_{4}^{-1}$) corresponding to the individual acoustic and optical phonon modes of unstrained and strained ML-ZnO at 300 K plotted as a function of the reduced wave vector along the k-path $\Gamma$-M.}
	\label{Fig.11}
\end{figure}

To get further insight into the 4-phonon scattering processes, we have calculated the mode-resolved $\tau_{4}^{-1}$ at 300 K for a particular k-path along $\Gamma$-M at different strain levels, as shown in Fig. \ref{Fig.11}. For pristine ML-ZnO, the relative strength of 4-phonon scattering ($\tau_{4}^{-1}$/$\tau_{3}^{-1}$) near the BZ center is found to be the highest for the ZA phonons. Such a large $\tau_{4}^{-1}$/$\tau_{3}^{-1}$ of the ZA phonons occurs due to the combined effect of (1) quadratic dispersion, (2) reflection symmetry selection rule (RSSR). The RSSR along with the large frequency gap in the phonon dispersion lead to an underestimation of the $\tau_{3}^{-1}$ associated with the ZA phonons. In a 4-phonon process, on the other hand, zero, two or four ZA phonons can involve, compared to the zero or two ZA phonons in a 3-phonon process. Thereby, 4-phonon processes such as, ZA+ZA = ZA+ZA or ZA = ZA+ZA+ZA are allowed. These additional scattering channels coupled with the large population of ZA phonons in the low-frequency region, resulting from the quadratic dispersion, gives rise to a strikingly high $\tau_{4}^{-1}$ close to the $\Gamma$-point. With the application of strain, owing to the quadratic to linear transition of the ZA mode dispersion, the phonon population decreases and thereby, the $\tau_{4}^{-1}$ drops. Unlike the $\tau_{3}^{-1}$, the $\tau_{4}^{-1}$ obtained at 4\% and 10\% strain are found to be significantly different, owing to the additional scattering channels allowed in 4-phonon scattering. For all the phonon modes, the $\tau_{4}^{-1}$ decreases with the application of strain with different magnitudes. However, the reduction in $\tau_{4}^{-1}$ is more significant for the acoustic modes compared to the optical modes. The variation in the $\tau^{-1}$, which is the sum of the $\tau_{3}^{-1}$ and $\tau_{4}^{-1}$, are computed at 300 K along the k-path $\Gamma$-M under different strained conditions, is shown in Fig. S4.

\begin{figure}[h!]
	\centering
	\includegraphics[scale=0.4]{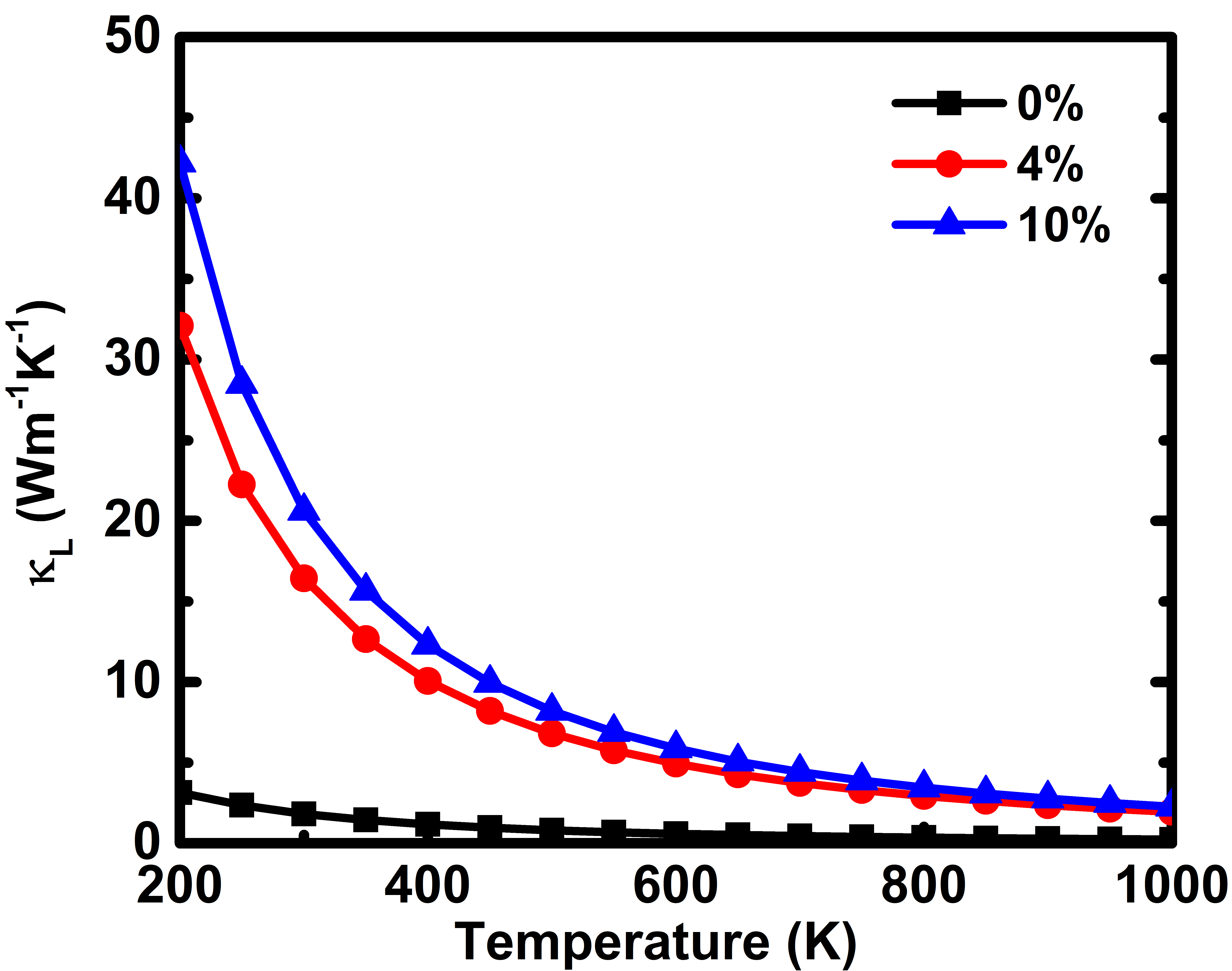}
	\caption{Variation in the thermal conductivity ($\kappa_{\text{L}}$) of ML-ZnO as a function of temperature under different tensile strains considering both the 3-phonon and 4-phonon scattering.}
	\label{Fig.12}
\end{figure}

\begin{table}[h]
	\caption{\label{tab:Table 3} Room temperature (300 K) Lattice thermal conductivity ($\kappa_{\text{L}}$) of ML-ZnO for the different strained conditions considering the 3-phonon scattering processes alone, and both the 3-phonon and 4-phonon scattering processes.}
	\centering
\begin{tabular*}{0.45\textwidth}{|c|@{\extracolsep{\fill}}c|@{\extracolsep{\fill}}c|}
	\hline 
	& \multicolumn{2}{c|}{$\kappa_{\text{L}}$ (Wm$^{-1}$K$^{-1}$)}\tabularnewline
	\hline 
	Strain (\%) & 3-phonon & 3-phonon + 4-phonon\tabularnewline
	\hline 
	0 & 6.1 & 1.8\tabularnewline
	\hline 
	4 & 45.6 & 16.4\tabularnewline
	\hline 
	10 & 54.5 & 20.6\tabularnewline
	\hline 
\end{tabular*}
\end{table}

\begin{figure}[h!]
	\centering
	\includegraphics[scale=0.35]{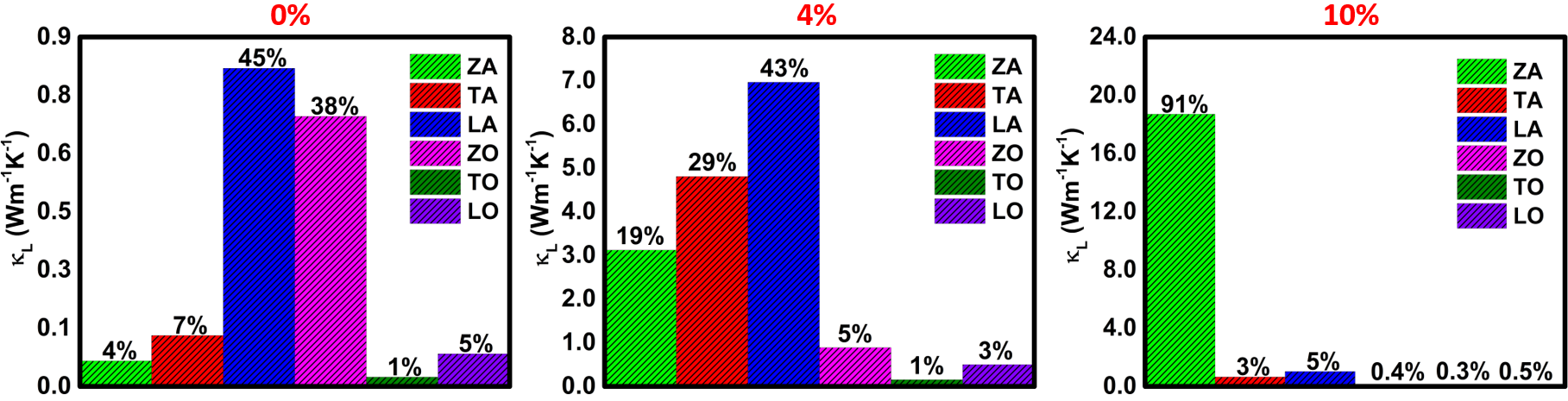}
	\caption{Percentage modal contribution to the total lattice  thermal conductivity ($\kappa_{\text{L}}$) stemming from the individual phonon modes at 300 K under the different strained conditions in ML-ZnO.}
	\label{Fig.13}
\end{figure}

Finally, combining the $\tau_{4}^{-1}$ along with the previously calculated $\tau_{3}^{-1}$, and the $\nu_{\lambda}$ and C$_\lambda$ extracted from the harmonic IFCs, we have computed the variation in $\kappa_{\text{L}}$ as a function of temperature for 0, 4 and 10\% strained cases, and shown in Fig. \ref{Fig.12}. The room temperature values of $\kappa_{\text{L}}$ of ML-ZnO, taking both the 3-phonon and 4-phonon scattering into account, are provided in Table \ref{tab:Table 3}. Similar increasing trends in $\kappa_{\text{L}}$ with strain, to what have been seen considering only the 3-phonon scattering, have been obtained even after the inclusion of higher-order anharmonicity. However, the quantitative aspects have been significantly modified while both the 3-phonon and 4-phonon scattering terms are considered. Due to the inclusion of 4-phonon scattering into the calculations and its unusually high strength, the $\kappa_{\text{L}}$ decreases throughout the temperature range of 200 to 1000 K for both the unstrained and strained cases. However, the influence of 4-phonon scattering decreases monotonically with increasing strain, as the reduction in $\kappa_{\text{L}}$ caused by the incorporation of $\tau_{4}^{-1}$ decreases from 71\% in the unstrained to 64\% in 4\% strained to 62\% in the 10\% strained condition. To understand the impact of 4-phonon scattering in greater detail, we have investigated the mode resolved $\kappa_{\text{L}}$ at 300 K for the different strained cases, and shown in Fig. \ref{Fig.13}. In the unstrained condition, owing to the large $\tau_{4}^{-1}$, the thermal transport contribution from the low-frequency ZA and TA phonons and high-frequency TO and LO phonons are significantly suppressed. Due to the comparatively lower $\tau_{4}^{-1}$ strength and a decent $\nu_{\lambda}$, the LA and ZO mode phonons carries the majority of the heat in the unstrained condition. With increasing strain, the contribution from the ZA mode grows, and at 10\% strain the ZA phonons govern almost the entire thermal transport characteristics of ML-ZnO. This dominance of the ZA mode with increasing strain can be understood as for the ZA phonons not only the $\tau^{-1}$ decreases with strain, but also the $\nu_{\lambda}$ increase significantly. These two effects act in unison to result in a dramatic enhancement in the thermal transport properties corresponding to the ZA phonons. The losing significance of the 4-phonon scattering is clear from the fact that, at higher values of strain, such as 10\%, the qualitative description on the role of the individual phonon modes in thermal transport obtained with or without the incorporation of 4-phonon scattering produces nearly identical results. It is, therefore, clear that, whether or not 4-phonon scattering is included, the intrinsic phonon scattering rate alone cannot explain the complex enhancement in $\kappa_{\text{L}}$ with the applied tensile strain. It is the competing behavior of $\nu_{\lambda}$ and $\tau_{\lambda}$, and their counter-balancing effects that lead to the observed anomalous variation in $\kappa_{\text{L}}$.

\section{Conclusions}
In summary, first-principles calculations have been performed to investigate the effects of in-plane isotropic biaxial tensile strains on the thermal transport properties of ML-ZnO. It is found that the lattice thermal conductivity ($\kappa_{\text{L}}$) of ML-ZnO increases significantly with the application of tensile strain, opposing the conventional wisdom. The strain induced quadratic to linear transition of the ZA mode dispersion not only results in an increase of the group velocity ($\nu_{\lambda}$), but also decreases the 3-phonon phase space (P$_{3}$) owing to the reduced phonon population. This, coupled with the dramatically reduced Gr$\ddot{\text{u}}$neisen parameter ($\gamma$), results in a large ZA phonon lifetime ($\tau$), which plays a major role behind the unusual enhancement of $\kappa_{\text{L}}$ with strain. The strain driven vying effects between the $\nu_{\lambda}$ and $\tau_{3}^{-1}$ lead to the up and down behavior in the thermal transport contribution of the other phonon modes. The $\tau$ of the ZA phonons are found to be brutally overestimated when only 3-phonon scattering is considered owing to the (1) quadratic dispersion, (2) large frequency gap in the phonon dispersion, and (3) reflection symmetry in ML-ZnO. The higher-order anharmonicity, albeit mainly the subsequent incorporation of the 4-phonon scattering, suppresses the $\tau$ and thereby, the $\kappa_{\text{L}}$ of the ZA phonons remarkably. However, the magnitude of reduction decreases with increasing strain, as the 4-phonon scattering loses strength at higher strain. The inclusion of 4-phonon scattering not only leads to a better accuracy in the prediction of $\kappa_{\text{L}}$, but also provide a much better understanding of the role of the individual phonon modes in the overall thermal transport. This work, therefore, provide a useful way to enhance the thermal transport properties of ML-ZnO along with a critical insight of the underlying mechanism. Furthermore, the investigation on the higher-order anharmonicity significantly advances the thermal transport studies of ML-ZnO or any 2D materials with analogous crystal structure.

\begin{acknowledgments}
The first-principles calculations have been performed using the supercomputing facility of IIT Kharagpur established under the National Supercomputing Mission (NSM), Government of India and supported by the Centre for Development of Advanced Computing (CDAC), Pune. AB acknowledges SERB POWER grant (SPG/2021/003874) and BRNS regular grant (BRNS/37098) for the financial assistance. SC acknowledges MHRD, India, for financial support.
\end{acknowledgments}
	
\bibliographystyle{achemso}
\bibliography{biblio}
	
\end{document}